\documentstyle[12pt,epsf]{article}

\input epsf

\newcommand{\be}{\begin{eqnarray}}

\newcommand{\ee}{\end{eqnarray}}

\newcommand{\nn}{\nonumber\\ }
\def\labe{\label}

\def\simge{\mathrel{%
   \rlap{\raise 0.511ex \hbox{$>$}}{\lower 0.511ex \hbox{$\sim$}}}}
\def\simle{\mathrel{
   \rlap{\raise 0.511ex \hbox{$<$}}{\lower 0.511ex \hbox{$\sim$}}}}
\def\bigs{\mathrel{
   \rlap{\raise 0.531ex \hbox{$>$}}{\lower 0.531ex \hbox{$<$}}}}

\def\grad{\nabla}                               % gradient
\def\del{\partial}                              % synonym for \partial

\def\frac#1#2{{#1 \over #2}}

\def\half{\ifinner {\scriptstyle {1 \over 2}}
   \else {1 \over 2} \fi}

\def\simge{\mathrel{%
   \rlap{\raise 0.511ex \hbox{$>$}}{\lower 0.511ex \hbox{$\sim$}}}}
\def\simle{\mathrel{
   \rlap{\raise 0.511ex \hbox{$<$}}{\lower 0.511ex \hbox{$\sim$}}}}
\def\bigs{\mathrel{
   \rlap{\raise 0.531ex \hbox{$>$}}{\lower 0.531ex \hbox{$<$}}}}

\def\slashchar#1{\setbox0=\hbox{$#1$}           % set a box for #1 
   \dimen0=\wd0                                 % and get its size
   \setbox1=\hbox{/} \dimen1=\wd1               % get size of /
   \ifdim\dimen0>\dimen1                        % #1 is bigger
      \rlap{\hbox to \dimen0{\hfil/\hfil}}      % so center / in box
      #1                                        % and print #1
   \else                                        % / is bigger
      \rlap{\hbox to \dimen1{\hfil$#1$\hfil}}   % so center #1
      /                                         % and print /
   \fi}                                         %

\def\subrightarrow#1{%                          % #1 under arrow
  \setbox0=\hbox{%                              % set a box
    $\displaystyle\mathop{}%                    % no mathop
    \limits_{#1}$}%                             % just limits
  \dimen0=\wd0%                                 % get width
  \advance \dimen0 by .5em%                     % add a bit
  \mathrel{%                                    % space like =
    \mathop{\hbox to \dimen0{\rightarrowfill}}% % arrow to width
       \limits_{#1}}}                           % text below

\begin{document}
\begin{titlepage}
%\vspace*{1.2cm}

\begin{flushright}
%BNL-NT-01/??\\
Saclay-T01/026\\hep-ph/0103032
\end{flushright}
\vspace*{1.2cm}
\begin{center}
{\Large{\bf Saturation and Universality in QCD at small $x$}}
\vskip0.3cm
 Edmond Iancu,\footnote{CNRS fellow. E-mail: eiancu@cea.fr}\\
{\small\it Service de Physique Th\'eorique\footnote{Laboratoire 
de la Direction des
Sciences de la Mati\`ere du Commissariat \`a l'Energie
Atomique}, CEA Saclay, 91191 Gif-sur-Yvette, France}
        
Larry McLerran\footnote{E-mail: mclerran@quark.phy.bnl.gov }\\
       {\small\it Physics Department, Brookhaven National Laboratory,
                 Upton, NY 11979, USA  }

\end{center}

\date{\today}
\vskip 1.2cm

\parindent=20pt

\begin{abstract}

We find approximate solutions to 
the renormalization group equation which governs
the quantum evolution of the effective theory for the 
Color Glass Condensate. This is a functional Fokker-Planck equation 
which generates in particular the non-linear evolution equations
previously derived by Balitsky and Kovchegov within perturbative QCD. 
In the limit where the transverse momentum of the
external probe is large compared to the saturation momentum, 
our approximations yield the Gaussian Ansatz 
for the effective action of the McLerran-Venugopalan model.  
In the opposite limit, of a small external momentum, we find that
the effective theory is governed by a scale-invariant universal action
which has the correct properties to describe gluon saturation.

\end{abstract}
\end{titlepage}

\setcounter{equation}{0}

There has been significant progress in recent years towards 
understanding the hadron structure in the high-density
regime at small $x$, which is the regime relevant for high-energy
scattering. As $x$ decreases, the gluon density in the
hadron wavefunction grows faster than the quark density
\cite{BFKL,HERA}, giving rise to
a multiparticle gluonic state with high occupation numbers.
In this high density environment, important non-linear phenomena are
expected, which should eventually lead to {\it saturation} 
\cite{GLR,MQ,MV94,AM2,GBW99}, that is, to a limitation of the maximum
gluon density per unit of phase-space. Linear evolution equations
like BFKL \cite{BFKL} do not take into account the rescattering
among the produced gluons, and predict an exponential growth of the
gluon distribution with $\ln(1/x)$. This leads to cross 
sections which in the high-energy limit violate the Froissart 
unitarity bound. If parton distributions saturate, then there is a
natural resolution to this unitarity problem.

Non-linear generalizations of the BFKL equation have been recently
proposed within different approaches, with results
which appear to be consistent with each other \cite{B,K,ILM}.
By using the operator expansion for high-energy scattering in QCD, 
Balitsky has derived a set of coupled equations for
the evolution of Wilson-line operators \cite{B}. 
This is an infinite hierarchy of coupled equations for
$n$-point operators, which
however decouple in the large $N$ limit, with $N$ the number of colors. 
In this limit, the equation satisfied by the 2-point function 
has been independently derived by Kovchegov
\cite{K} within the dipole model of Mueller \cite{AM3}. 
Solutions to Kovchegov's equation 
have been investigated in Refs. \cite{K,LT99,B00}.
In a recent paper \cite{W}, Weigert has 
shown that Balitsky's equations can be summarized as a functional 
evolution equation for the generating functional of the Wilson line 
operators. Together with Leonidov, we have subsequently shown \cite{ILM}
that this is precisely the renormalization group equation (RGE)
which governs the quantum evolution of the effective theory for
the Color Glass Condensate (CGC) \cite{JKLW97,PI}.

In the description of saturation, a natural intrinsic 
momentum scale appears, the saturation momentum $Q_{s}$
\cite{MV94,AM2,JKMW97,KM98,AM1}. This is the scale below which
the non-linear effects are expected to slow down and eventually
saturate the increase of the gluon density with
$1/k_\perp^2$, where $k_\perp$ is the gluon transverse 
momentum\footnote{Throughout, we consider
the hadron in its infinite momentum frame, and use light-cone vector 
notations,  $v^\mu=(v^+,v^-,{\bf v}_\perp)$, with
$v^+\equiv (1/\sqrt 2)(v^0+v^3)$,
$v^-\equiv (1/\sqrt 2)(v^0-v^3)$, and ${\bf v}_\perp
\equiv (v^1,v^2)$. In these notations, the hadron four-momentum
reads $P^\mu=(P^+,0,0_\perp)$, with $P^+\to \infty$.}.
In this paper, we shall find approximate solutions to
the RGE for the CGC in two limits: at large momenta $k_\perp \gg Q_{s}$,
and at small momenta $k_\perp \ll Q_{s}$, where $k_\perp$
is the momentum scale at which we compute correlation functions.
($a$) In the first case,
our results reproduce the Gaussian Ansatz for the ``effective action''
(strictly speaking, a probability distribution for a classical
color source; see below) which has been 
postulated in the McLerran-Venugopalan (MV) model \cite{MV94}. 
We shall find improvements over the previous discussions of this model
\cite{MV94,JKMW97,KM98}
in the sense of including the quantum evolution of the width of the 
Gaussian and properly treating the transverse non-locality.
We shall also clarify the limits of validity of this model, 
and show that it does not apply at low momenta $k_\perp \simle Q_{s}$,
which is the interesting regime for saturation. ($b$) In the second
case, at $k_\perp \ll Q_{s}$, our approximate solution is
still a Gaussian, but with a different structure.
{ It is scale-invariant and universal (up to logarithmic corrections
which enter via the matching with the solution in the high-momentum
regime), and has the right properties to describe gluon saturation.
The associated gluon density increases only logarithmically with
$1/k_\perp^2$, a result formally similar to that obtained in
the classical MV model \cite{JKMW97,KM98},
but which here arises in a quite different way: The non-linear
effects leading to saturation are encoded in the quantum
evolution, and not in the solution to the classical equations
of motion. Our results appear to be consistent with other
recent analyses of the non-linear gluon evolution \cite{AM2,LT99}.}

In the MV model \cite{MV94}, gluon correlation 
functions at small $x$ are obtained as classical correlations
within the stochastic Yang-Mills theory with the
following equation of motion
\be
(D_{\nu} F^{\nu \mu})_a(x)\, =\, \delta^{\mu +} \rho_a(x)\,,
\labe{cleq0}
\ee
where $\rho_a(x)$ is the effective color charge density
at the ``soft'' scale $k^+ \equiv xP^+$ due to the ``fast''
partons with longitudinal momenta $p^+ \gg k^+$.
Because of this separation of scales, the source is 
time-independent and localized near the 
light-cone, within a distance $\Delta x^- \simle 1/k^+$.
It is furthermore a random quantity whose spatial correlators
can be summarized as a gauge-invariant weight function 
$W_\tau[\rho]$, which is the probability for having 
a color charge distribution with density $\rho_a(\vec x)$
(with ${\vec x}\equiv (x^-,{\bf x}_{\perp})$).
We have introduced here the momentum-space rapidity 
$\tau\equiv\ln(P^+/k^+) = \ln(1/x)$ to indicate the
dependence of the weight function upon the soft scale $k^+$.

Thus, equal-time gluon correlation functions at the scale $k^+ =
 xP^+$ are obtained as:
\be\labe{clascorr}
\langle A^i_a(x^+,\vec x)A^j_b(x^+,\vec y)
\cdots\rangle_\tau\,=\,
\int {\cal D}\rho\,\,W_\tau[\rho]\,{\cal A}_a^i({\vec x})
{\cal A}_b^j({\vec y})\cdots\,,\ee
where ${\cal A}_a^i\equiv {\cal A}_a^i[\rho]$ is the 
solution to eq.~(\ref{cleq0}) in the light-cone (LC) 
gauge $A^+_a=0$, which is the gauge which allows for the most
direct contact with the gauge-invariant physical quantities 
\cite{AM1,PI}. For instance, the gluon distribution function is
obtained as \cite{AM1,PI}
\be\labe{GCL}
x G(x,Q^2)&=&\frac{1}{\pi}
\int {d^2k_\perp \over (2 \pi)^2}\,\Theta(Q^2-
k_\perp^2)\,\Bigl\langle\,
|{\cal F}^{+i}_a(\vec k)|^2\Bigr\rangle_\tau\,,\ee
where ${\cal F}^{+i}_a=\partial^+ {\cal A}^i_a$ is the 
classical electric field, and $\vec k \equiv (k^+,{\bf k}_\perp)$
with $k^+=xP^+=P^+{\rm e}^{-\tau}$. However, to have
explicit expressions for these classical fields,
it is preferable to express the LC-gauge solution
${\cal A}_a^i[\rho]$ in terms of color source $\rho$
in the covariant gauge $\partial^\mu \tilde A_\mu =0$
(COV-gauge). 
One then obtains (with $\rho\equiv \rho_a T^a$, etc.)\cite{PI}:
\be\labe{Atilde}\,
{\cal A}^i\,
(\vec x) &=&{i \over g}\, U(\vec x) \,\partial^i  U^\dagger(\vec x),\nn
U^{\dagger}(x^-,x_{\perp})&=&
 {\rm P} \exp
 \left \{
ig \int_{-\infty}^{x^-} dz^-\,{\alpha}(z^-,x_{\perp})
 \right \},\nn
- \nabla^2_\perp \alpha({\vec x})&=&\rho(\vec x),
\ee
where $\rho_a$, $\tilde{\cal A}^\mu_a =\delta^{\mu +}
\alpha_a$, and $\tilde {\cal F}^{+i}_a=-\partial^i\alpha_a$
are respectively the color source, the vector potential, and
the electric field in the COV-gauge. 
In the saturation regime,
the source and the fields are strong,
${\cal A}^i\sim \rho \sim 1/g$, 
and the classical problem is fully non-linear.

The source $\rho_a$ and the associated
weight function $W_\tau[\rho]$ are constructed in perturbation theory,
by integrating out the fast quantum gluons in layers of $p^+$ to
leading order in $\alpha_s\ln(1/x)$, but
to all orders in the strong background fields ${\cal A}^i$
generated at the previous steps \cite{JKMW97,JKLW97,PI}.
With the gauge-fixing prescriptions advocated in Refs. \cite{ILM,PI},
this implies that the source $\rho_a(\vec x)$
has support only at positive $x^-$, with $x^-_0\simle x^-\simle 1/k^+$,
where $x^-_0 \equiv 1/P^+$ is the Lorentz-contracted 
longitudinal size of the hadron.
By introducing the space-time rapidity ${\rm y}\equiv \ln(x^-/x^-_0)$
and recalling that $1/k^+ = 1/(xP^+) = x^-_0{\rm e}^\tau$, we conclude
that $\rho_a$ has support at $0\le {\rm y}\le \tau$.

The evolution of the weight function $W_\tau[\rho]$
with increasing $\tau=\ln(1/x)$ is described by a functional
renormalization group equation
 originally proposed by Jalilian-Marian, Kovner, Leonidov and Weigert 
\cite{JKLW97} and explicitely constructed in Refs.
\cite{PI,ILM} (see also Refs. \cite{JKW99,KMW00}). 
It acquires its simplest form when written as
an equation for $W_\tau[\alpha]$ (recall that $\alpha$ and $\rho$
are linearly related, cf. eq.~(\ref{Atilde})), in which case
it reads \cite{ILM}
\be\labe{RGEH}
{\del W_\tau[\alpha] \over {\del \tau}}&=&-\,H W_\tau, 
\ee
with the following, positive definite, Hamiltonian:
\be\labe{H}
H&=&\int {d^2 z_\perp\over 2\pi }\,J^i_a(z_\perp)\,J^i_a(z_\perp),\nn
J^i_a(z_\perp)&\equiv& \int {d^2 x_\perp\over 2\pi }\,
\frac{z^i-x^i}{(z_\perp-x_\perp)^2}\,(1 - V^\dagger_zV_x)_{ab}\,
{i \delta \over {\delta
\alpha_\tau^b(x_\perp)} }.\ee
This is the same Hamiltonian which generates Balitsky's evolution
 equations, as rewritten in functional form
by Weigert \cite{W}; this establishes the equivalence between
the approaches of Refs. \cite{B} and \cite{ILM,PI}. 

In the equations above,
\be
V^\dagger_x\,\equiv\,V^\dagger(x_\perp)
\,\equiv\,U^\dagger(x^-=x^-_0{\rm e}^\tau,x_\perp),\qquad
\alpha_\tau^a(x_\perp)\,\equiv\,\alpha^a(x^-=x^-_0{\rm e}^\tau,x_\perp),
\ee 
are the Wilson line and the COV-gauge ``Coulomb'' field in
eq.~(\ref{Atilde}) evaluated at $x^-=x^-_0{\rm e}^\tau=1/k^+$,
which is the largest longitudinal coordinate for the color source
at the scale $k^+=P^+{\rm e}^{-\tau}$. 
Thus, the functional derivatives in eqs.~(\ref{RGEH})--(\ref{H})
are to be taken with respect to the color field 
in the highest bin of (space-time)
rapidity $\tau \le {\rm y}\le \tau+d\tau$,
where the quantum corrections are located.
This suggests a space-time picture of the quantum evolution
where the momentum rapidity $\tau$ and the space-time rapidity
y are interrelated: The classical source 
\be\rho_{\rm y}^a(x_\perp)
\equiv \rho^a(x^-=x^-_0{\rm e}^{\rm y},x_\perp)\ee
is constructed
in layers of space-time rapidity, with the contribution in the
rapidity bin $({\rm y},\, {\rm y} +d{\rm y})$ obtained by integrating
out the quantum gluons with longitudinal momenta in the
momentum-rapidity bin $(\tau,\,\tau+d\tau)$ with $\tau={\rm y}$.
This relation will become explicit in the solutions to the RGE 
to be found below.

Eq.~(\ref{RGEH}) is a functional Fokker-Planck equation with ``time''
$\tau$. It depicts the quantum evolution towards small $x$ as the 
diffusion of the probability density $W_\tau[\alpha]$
in the functional space spanned by $\alpha_a(x^-,x_\perp)$.
Since the r.h.s. of this equation can be written as
a total derivative with respect to $\alpha$ \cite{ILM,PI}, 
the correct normalization:
\be\label{norm}
\int {\cal D}\alpha\, \,W_\tau[\alpha]\,=\,1\,,\ee
is automatically preserved by this evolution. In the weak field
limit, where the Wilson lines can be expanded to lowest order in
$\alpha$, eqs.~(\ref{RGEH})--(\ref{H}) reproduce
the BFKL equation for the charge-charge
correlator $\langle\rho(k_\perp)\rho(-k_\perp)\rangle$ \cite{JKLW97,PI}.
In the strong field regime $\alpha\sim 1/g$ specific to saturation,
eq.~(\ref{RGEH}) is equivalent to an infinite hierarchy of coupled
evolution equations for the correlators of $\alpha$.
In what follows, we shall not study these coupled equations, but
rather attempt to solve directly the functional RGE (\ref{RGEH})
in some limiting cases. More details will be presented
somewhere else \cite{SATL}.

As announced in the introduction, we shall find approximate
solutions to  eq.~(\ref{RGEH}) in two different 
kinematical regimes: $k_\perp\gg Q_s(\tau)$ and $k_\perp\ll Q_s(\tau)$,
where $k_\perp$ is the transverse momentum at which we measure
the correlation functions (i.e., the transverse resolution
of the external probe),
while $Q_s(\tau)$ (``the saturation momentum'') is an intrinsic
momentum scale which is introduced by the initial
conditions (note that there is no explicit scale in the evolution
equation (\ref{RGEH})--(\ref{H})) and which increases with $\tau$.
No attempt will be made to describe the intermediate behaviour 
at $k_\perp \sim Q_s$, and thus the onset of saturation.
Because of that, the saturation scale itself will be not exactly
determined, but only estimated via a study of the onset of
non-linearity in the high momentum regime.

{\bf a) The high momentum regime.}
As we shall shortly discover, at least within the approximations
to be considered here, it is the same scale $Q_s$ which controls
both the transverse correlation length in the problem,
and the onset of the non-linear regime. That is, 
the typical variation scale for the Wilson line $V(x_\perp)$ is
$1/Q_s$, and the non-linear effects become
important at transverse momenta of order $Q_s$ or less.
Thus, as long as we are probing the system on a much shorter resolution
scale, corresponding to some external momentum $k_\perp\gg Q_s$,
we can neglect the non-linear effects and perform 
a short-distance expansion in the Hamiltonian (\ref{H}) :
\be\label{APPROXHM}
1 - V^\dagger_zV_x\,\approx\, (z^j-x^j)(\partial^j V^\dagger)_x V_x
\,\approx\,ig(z^j-x^j)\partial^j\alpha(x_\perp),\ee
where (with $x^-_\tau\equiv x^-_0{\rm e}^\tau$)
\be
\alpha^a(x_\perp)\,\equiv\,
\int_{x^-_0}^{x^-_\tau} dx^- \alpha^a(x^-,x_\perp) \,=\,
\int_0^\tau d{\rm y}\,\hat\alpha_{\rm y}^{\, a}(x_\perp),\ee
is the effective Coulomb field in the transverse plane,
and we have introduced, for further convenience, the notation
$\hat\alpha_{\rm y}^{\, a}(x_\perp)\equiv x^-\alpha^a(x^-,x_\perp)$
for $x^-=x^-_0{\rm e}^{\rm y}$.

With these approximations, and after simple manipulations to be
detailed in Ref. \cite{SATL}, the Hamiltonian (\ref{H}) is brought
into the following form:
\be\label{HHM}
H\,\approx\,{g^2\over 2\pi }\int {d^2 x_\perp}
\int {d^2 y_\perp}\,\langle x_\perp|\frac{1}{\grad^4_\perp}|y_\perp\rangle\,
\partial^i\alpha^{ab}(x_\perp)\,{ \delta \over {\delta
\alpha_\tau^b(x_\perp)}}\,\partial^i\alpha^{ac}(y_\perp)\,
{\delta \over {\delta \alpha_\tau^c(y_\perp)} }\,.\ee
The RGE associated to this Hamiltonian is still non-linear, and
we have not been able to solve it exactly.
To make progress, we perform a mean field approximation
in which we replace, within eq.~(\ref{HHM}),
\be\label{Edef}
\partial^i\alpha^{ab}(x_\perp)\,
\partial^i\alpha^{ac}(y_\perp)
\,\rightarrow\,\langle \partial^i\alpha^{ab}(x_\perp)\,
\partial^i\alpha^{ac}(y_\perp)\rangle_\tau\,\equiv\,
N\delta^{bc}\grad^2_x\xi_\tau(x_\perp-y_\perp),
\ee
where the expectation value in the r.h.s. is defined 
as in eq.~(\ref{clascorr}), and we have assumed homogeneity in the
transverse plane, for simplicity. (The
color structure in the r.h.s. follows from gauge invariance.)
With this approximation, the RGE is linear and can be solved
by a Gaussian weight function. The correlation function
$\xi_\tau$ in eq.~(\ref{Edef}) can be then
computed in terms of the width of this Gaussian, which,
as we shall see, entails a self-consistency condition describing
the evolution of the width with $\tau$.

Specifically, we have to solve the following equation:
\be\labe{RGEHM}
{\del W_\tau[\alpha] \over {\del \tau}}&=&-{1\over 2}
\int {d^2 x_\perp}\int {d^2 y_\perp}\,{\cal D}_\tau(x_\perp-y_\perp)\,
\frac{\delta^2W_\tau[\alpha]}{\delta \alpha_\tau^a(x_\perp)
\delta \alpha_\tau^a(y_\perp)},\nn
{\cal D}_\tau(x_\perp-y_\perp)&\equiv&{g^2N\over \pi}\,
\langle x_\perp|\frac{1}{\grad^4_\perp}|y_\perp\rangle\,
{\xi}_\tau(x_\perp-y_\perp),
\ee
or, in momentum space,
\be
{\cal D}_\tau(k_\perp)\,=\,{g^2N\over \pi}
\int{d^2p_\perp\over (2\pi)^2}\,{p_\perp^2\over(k_\perp-p_\perp)^4}\,
{\xi}_\tau(p_\perp)\,\simeq\,{g^2N\over \pi}\,{1\over k_\perp^4}\,
\int^{k_\perp}{d^2p_\perp\over (2\pi)^2}\,p_\perp^2{\xi}_\tau(p_\perp)\,,\ee
where in writing the approximate equality we have used the fact
that $k_\perp$ is the hard external momentum, 
while $p_\perp$ is relatively soft,
$p_\perp\sim Q_s \ll k_\perp$. Since
$k_\perp^2\alpha_\tau(k_\perp)=\rho_\tau(k_\perp)$, cf. eq.~(\ref{Atilde}),
the Hamiltonian in eq.~(\ref{RGEHM}) is finally rewritten as
(with $\alpha_s=g^2/4\pi$):
\be\label{HHMR}
H_{{\rm high}-k_\perp}&=&-\,{1\over 2}\int{d^2k_\perp\over (2\pi)^2}\,
{\cal G}_\tau(k_\perp)\,
\frac{\delta^2}{\delta \rho_\tau^a(k_\perp)
\delta \rho_\tau^a(-k_\perp)},\nn
{\cal G}_\tau(k_\perp)&\equiv& 4\alpha_s N
\int^{k_\perp}{d^2p_\perp\over (2\pi)^2}\,
p_\perp^2{\xi}_\tau(p_\perp)\,.\ee
Thus, in this high-momentum regime, the solution
to the RGE is most naturally written as a Gaussian in $\rho_a({\vec x})$.
The only subtle point about this solution refers to
its longitudinal structure. 
%[recall that $\rho_\tau(k_\perp)\equiv \rho(x^-_\tau,k_\perp)$
%with $x^-_\tau\equiv x^-_0{\rm e}^\tau$], the solution to this RGE
To understand this structure, note that the two functional derivatives
in eq.~(\ref{HHMR}) act at the same point $x^-$, namely at 
$x^-=x^-_\tau\equiv x^-_0{\rm e}^\tau$.
This implies that the
correlations generated by the RGE are local in $x^-$. We
thus search for a solution $W_\tau[\rho]$ of the form\footnote{The initial
condition for the solution will be discussed in Ref. \cite{SATL}.}:
\be\label{WHM}
W_\tau[\rho]\,=\,{\cal N}_\tau\,{\rm exp}\left\{
-\,{1\over 2}\int_0^\infty d{\rm y}
\int{d^2k_\perp\over (2\pi)^2}\,\hat\rho_{\rm y}^{\,a}(k_\perp)
\lambda^{-1}_\tau({\rm y},k_\perp) \hat\rho_{\rm y}^{\,a}(-k_\perp)
\right\},\ee
where we have  used the space-time rapidity y to indicate
the $x^-$--dependence of the various functions and defined
$\hat\rho_{\rm y}^{\,a}(x_\perp)\equiv x^-\rho^a(x^-,x_\perp)$
for $x^- = x^-_0{\rm e}^{\rm y}$.
The integration over y in eq.~(\ref{WHM}) runs effectively 
up to $\tau$, but the upper integration limit is conveniently
absorbed in the support of the kernel 
$\lambda^{-1}_\tau({\rm y},k_\perp)$ (see eq.~(\ref{LHM}) below).
The overall factor ${\cal N}_\tau$ follows from the
normalization condition (\ref{norm}) as (up to some irrelevant,
$\tau$-independent, factor):
\be
{\cal N}_\tau\,=\,[{\rm det} \,\lambda_\tau({\rm y},k_\perp)]^{-1/2}
\,=\,\prod_{k_\perp}\,\prod_{\rm y} 
\,[\lambda_\tau({\rm y},k_\perp)]^{-(N^2-1)/2},
\ee
where in writing
the second equality we have considered a lattice
version of the 3-dimensional configuration space, with discrete points 
$k_\perp$ and y. 

It is now straightforward to verify that the functional (\ref{WHM})
satisfies the RGE associated to
$H_{{\rm high}-k_\perp}$ provided the width 
$\lambda_\tau({\rm y},k_\perp)$ obeys the following equation: 
\be\label{SCEHM}
\frac{\partial \lambda_\tau({\rm y},k_\perp)}{\partial\tau}\,=\,
\delta(\tau-{\rm y}){\cal G}_\tau(k_\perp),\ee
which shows that the evolution of the width with the
{\it momentum-space} rapidity $\tau$ takes place at
the {\it space-time} rapidity y$\,\,=\tau$,
that is, at the end point of the
color charge distribution at the previous step.
The solution to eq.~(\ref{SCEHM}) is immediate:
\be\label{LHM}
\lambda_\tau({\rm y},k_\perp)\,=\,\theta(\tau-{\rm y})\,
{\cal G}_{\rm y}(k_\perp).\ee
This gives the following 2-point function for the color charge:
\be\label{rhoHM}
\langle \hat\rho_{\rm y}^{\,a}(x_\perp)\,
\hat\rho_{{\rm y}'}^{\,b}(y_\perp)\rangle_\tau\,=\,\delta^{ab}\delta({\rm y}
-{\rm y}')\theta(\tau-{\rm y}){\cal G}_{\rm y}(x_\perp-y_\perp).\ee
{ Note that, for ${\rm y} < \tau$,
the correlation function ${\cal G}_{\rm y}(x_\perp-y_\perp)$
is independent of $\tau$, since determined uniquely
 by the quantum evolution up to the momentum rapidity y
 (via eq.~(\ref{delmu}) below).
%This reflects the fact that the color charge distribution is
%constructed by independently adding contributions in successive
%layers of rapidity.

According to eqs.~(\ref{WHM}) and (\ref{LHM}), the width of the
Gaussian is the {\it inverse} of a $\theta$-function, which 
means that the weight function $W_\tau[\rho]$ 
is identically zero for all the functions $\rho^a_{\rm y}(x_\perp)$ 
having support at rapidities ${\rm y} > \tau$. Thus, in the functional
integral (\ref{clascorr}) one can freely integrate over all the functions
$\rho^a_{\rm y}(x_\perp)$, without any restriction on their support
(other than ${\rm y} > 0$); the restriction to ${\rm y} < \tau$ will
be automatically taken care of by the weight function.
}

At this point, one should recall that the correlation function
${\xi}_\tau$, eq.~(\ref{Edef}), and thus the function
${\cal G}_\tau(k_\perp)$ in eq.~(\ref{HHMR}), are
 expectation values with the weight function (\ref{WHM}),
and thus functionals of $\lambda_\tau$.
Thus, eqs.~(\ref{SCEHM}) or (\ref{LHM}) are really 
self-consistency conditions, to be made explicit now. 
To this aim, it is convenient to integrate eq.~(\ref{SCEHM})
over ${\rm y}$,
\be\label{delmuG}
{\partial \mu_\tau(k_\perp^2)\over \partial \tau}\,=\,
{\cal G}_\tau(k_\perp),\ee
and thus obtain an evolution equation for the quantity
\be\label{mu}
\mu_\tau(k_\perp^2)\,\equiv\,\int d{\rm y}\,
\lambda_\tau({\rm y},k_\perp)\,,\ee
which, in the present weak field regime, is the unintegrated gluon
distribution. To see this, note that, to lowest order in $\rho$, 
${\cal F}^{+j}_a \simeq i(k^j/k^2_\perp)\rho_a\,$,  so eq.~(\ref{GCL})
becomes
\be\labe{GCLLIN}
x G(x,Q^2)\,\simeq\,\frac{1}{\pi}
\int^Q {d^2k_\perp \over (2 \pi)^2}\,
\frac{\Theta(Q^2-k_\perp^2)}{k^2_\perp}\,\Bigl
\langle |\,\rho_a(\vec k)|^2\Bigr\rangle_\tau\,=\,
\frac{(N^2-1)R^2}{4\pi}\int_0^{Q^2}\frac{dk_\perp^2}{k_\perp^2}\,
\mu_\tau(k_\perp^2),\ee 
where  $R$ is the hadron radius, and
we have also used eqs.~(\ref{rhoHM}) and (\ref{mu}).

From eqs.~(\ref{Edef}) and (\ref{rhoHM})--(\ref{mu}), one obtains
%\be\label{ximu}
${\xi}_\tau(p_\perp)=(1/p_\perp^4)\mu_\tau(p_\perp^2)$, %\,,\ee
which, together with eqs.~(\ref{HHMR}) and (\ref{delmuG}),
provides an evolution equation for $\mu_\tau(k_\perp^2)\,$:
\be\label{delmu}
{\partial \mu_\tau(k_\perp^2)\over \partial \tau}\,=\,\frac{\alpha_s N}
{\pi} \int_0^{k_\perp^2}\frac{dp_\perp^2}{p_\perp^2}\,
\mu_\tau(p_\perp^2)\,.\ee 
By also using eq.~(\ref{GCLLIN}), this can be recognized as
 the evolution equation for
the gluon distribution function in the double logarithmic approximation
(DLA) \cite{DGLAP} :
\be\label{DL}
{\partial^2\over \partial \tau\,\partial\ln Q^2
}\,xG(x,Q^2)\,
=\,\frac{\alpha_s N}{\pi}\,xG(x,Q^2).\ee
Given the approximations that we have performed,
this is indeed the expected limit of the evolution equation \cite{SATL}. 
The solution to eq.~(\ref{DL}) is well known \cite{DGLAP} : 
if one holds $\alpha_s$ fixed (independent of $Q^2$),
the solution grows like:
\be\label{DLA}
xG(x,Q^2)\,\propto\,{\rm exp}\left\{2\sqrt{\frac{\alpha_s N}{\pi}\,
\tau\,\ln(Q^2/Q^2_0)}\right\}.\ee
%while by taking $\alpha_s(Q^2)\propto 1/\ln(Q^2/\Lambda^2_{QCD})$,
%the growth of $G$ with $Q^2$ is even softer (doubly logarithmic).
For our purposes here, what really matters is that the initial
condition to eq.~(\ref{delmu}) introduces a momentum scale
in the problem, while there was no such a scale in the RGE by itself. 
For instance, in a simple valence-quark model,
which should be a reasonable approximation
near $x\sim 1$  (or $\tau\sim 0$),
one has \cite{AM1} (with $C_F=(N^2-1)/2N$)
\be
x G(x,Q^2)\,\simeq\,\frac{\alpha_s N\, C_F}{\pi}\,
\ln{Q^2\over \Lambda^2_{QCD}}
%\,\equiv\,
%\frac{(N^2-1)R^2}{4\pi}\int_0^{Q^2}\frac{dk_\perp^2}{k_\perp^2}\,
%\mu_0(k_\perp^2),\ee 
\quad{\rm for}\,\,\,\,x\sim 1\,,\ee
where $\Lambda^2_{QCD}$ enters naturally as an infrared cutoff
when taking into account the over all color neutrality of the hadron
on scale sizes of order $1/\Lambda_{QCD}$ \cite{LM00}.
Thus, for $\tau\sim 0$,
\be
\mu_0(k_\perp^2)\,\sim\,2\alpha_s/R^2,\ee
which for a proton is a relatively small momentum scale to start
with, but which at low $x$ is enhanced by the quantum evolution described
by eq.~(\ref{delmu}).

When $\tau$, and therefore $\mu_\tau$, are sufficiently large, 
the fluctuations of the color field described by
eq.~(\ref{WHM}) become large as well,
and the non-linear effects start to play a role. To study the onset of
non-linearity, and verify the approximations performed
in eq.~(\ref{APPROXHM}),
it is useful to consider the 2-point function 
$\langle V^\dagger_x V_y\rangle_\tau$. This is easily computed by 
expanding the path-ordered
exponentials within the Wilson lines, performing the
contractions of the fields $\alpha$  with the help of
(cf. eqs.~(\ref{rhoHM}) and  (\ref{delmuG}))  %and (\ref{ximu}))
\be\label{alphaHM}
\langle \hat\alpha_{\rm y}^{\,a}(x_\perp)\,
\hat\alpha_{{\rm y}'}^{\,b}(y_\perp)\rangle_\tau&=&\delta^{ab}\delta({\rm y}
-{\rm y}')\theta(\tau-{\rm y})\,\gamma_{\rm y}(x_\perp-y_\perp),\\
\gamma_{\tau}(p_\perp)&=&{1\over p_\perp^4}\,
{\partial \mu_\tau(p_\perp^2)\over \partial \tau}
\,=\,{\partial \xi_\tau(p_\perp) \over \partial \tau}\,,\ee
and then recognizing the result as the expansion of an ordinary
exponential. (See Refs.
\cite{JKMW97,KM98} for similar calculations.)
The final result can be written as:
\be\label{VV}
\langle V^\dagger(x_\perp) V(0_\perp)\rangle_\tau=
{\rm exp}\left\{-g^2 N\int_0^\tau d{\rm y}\Bigl[\gamma_{\rm y}
(0_\perp)-\gamma_{\rm y}(x_\perp)\Bigr]\right\}=
{\rm e}^{-g^2 N[\xi_\tau(0_\perp)-\xi_\tau(x_\perp)]}\,,\ee
with
\be
\xi_\tau(0_\perp)-\xi_\tau(x_\perp)\,=\,\int 
{d^2p_\perp\over (2\pi)^2}\,
\frac{\mu_\tau(p_\perp^2)}{p_\perp^4}\,\Bigl[1-
{\rm e}^{ip_\perp\cdot x_\perp}\Bigr]\,.\label{xiox}\ee
Since $\mu_\tau(p_\perp^2)$ is rather
slowly varying as a function of $p_\perp$, the integral in 
eq.~(\ref{xiox}) is dominated by small momenta $p_\perp^2\ll
1/x_\perp^2$, where we can approximate
\be\label{xi1}
\int {d^2p_\perp\over (2\pi)^2}\,
\frac{1}{p_\perp^4}\,\Bigl[1-
{\rm e}^{ip_\perp\cdot x_\perp}\Bigr]\,\simeq\,
\int^{1/x_\perp^2} {d^2p_\perp\over (2\pi)^2}\,
\frac{1}{p_\perp^4}\,{(p_\perp\cdot x_\perp)^2\over 2}
\,\simeq\,{x_\perp^2\over 16\pi}\,
\ln\,{1\over x_\perp^2\Lambda^2_{QCD}}\,.\ee
%(This is valid to leading logarithmic accuracy, since the
%corresponding contribution of the hard momenta, $p_\perp^2\gg
%1/x_\perp^2$, is not enhanced by a large logarithm.)
By performing a similar approximation on eq.~(\ref{xiox}),
one obtains:
\be\label{xioxAP}
\langle V^\dagger(x_\perp) V(0_\perp)\rangle_\tau\,\simeq\,
{\rm exp}\left\{-\frac{\alpha_s N}{4}\,x_\perp^2 \int^{1/x_\perp^2}
\frac{dp_\perp^2}{p_\perp^2}\,
\mu_\tau(p_\perp^2)\right\}\,.\ee 
By inspection of this equation, it should be clear
that, as anticipated, it is the same
scale $Q_s^2\sim 1/x_\perp^2$ which controls both the non-locality
and the onset on non-linearity: this is
the scale for which the exponent in eq.~(\ref{xioxAP}) is of
order 1. This implies the following estimate (actually,
an equation) for the saturation scale:
\be\label{Qsat}
Q^2_s(\tau)\,\simeq\,\frac{\alpha_s N}{4}\,
\int^{Q^2_s}\frac{dp_\perp^2}{p_\perp^2}\,\mu_\tau(p_\perp^2)\,.\ee 
For $k_\perp \gg Q_s$, %(i.e., $x_\perp^2 \ll 1/Q^2_s$),
the approximations in eq.~(\ref{APPROXHM}) are justified,
and the Gaussian (\ref{WHM}) is the correct solution to the RGE
within the mean field approximation. { But this solution cannot be
extended at low momenta $k_\perp \simle Q_s$. This shows the
inconsistency of some previous calculations within the classical
MV model, where the Gaussian weight function (\ref{WHM})
has been used at {\it all} momenta, including in the saturation
regime \cite{JKMW97,KM98}.}

{ For later reference, it is useful to combine eqs.~(\ref{Qsat}),
(\ref{GCLLIN}) and
(\ref{DLA}) to obtain an estimate for the $\tau$-dependence of $Q_s$
in the DLA (with $\bar\alpha_s \equiv \alpha_s N/\pi$) :
\be\label{Qstau}
Q^2_s(\tau)\,\propto\,xG(x,Q^2_s)\,
\sim\,{\rm exp}\left\{2\sqrt{\bar\alpha_s
\tau\ln(Q^2_s/Q^2_0)}\right\},\qquad{\rm or}\quad
Q^2_s(\tau)\,\propto\,{\rm e}^{\,4\bar\alpha_s\tau}\,.
\ee
%A priori, this estimate is rather crude, since it relies on the
%high-momentum expansion. But a better estimate would require
%the solution to the RGE in the non-linear regime at $k_\perp \sim Q_s$.

}

{\bf b) The low momentum regime.}
We now turn to the more interesting regime at small momenta,
$k_\perp \ll Q_s(\tau)$, when the hadron is probed 
over distances large compared to
the correlation length $1/Q_s(\tau)$ (but still small as
compared to $1/\Lambda_{QCD}$). In addition to being relatively
rapidly varying, the electric fields in this regime
are also expected to have large amplitudes, of order $1/g$
(see eq.~(\ref{WIGG}) below).
In this regime, operators like $V^\dagger_z V_x$ are rapidly 
averaging to zero. Thus, in a first approximation, we shall 
simply ignore all the Wilson lines in the Hamiltonian (\ref{H}).
We shall verify a posteriori that this is a consistent
approximation, by computing the expectation value
$\langle V^\dagger_x V_y\rangle_\tau$ with the resulting weight function.
This approximation will be further justified in \cite{SATL}.

In this ``random phase
approximation'' (RPA), the Hamiltonian reads simply:
\be\label{HLM}
H_{{\rm low}-k_\perp}\,\approx\,-\,{1\over 2\pi }\int {d^2 x_\perp}
\int {d^2 y_\perp}\,\langle x_\perp|\frac{1}{-\grad^2_\perp}|y_\perp\rangle\,
{ \delta^2 \over {\delta \alpha_\tau^a(x_\perp)
\delta \alpha_\tau^a(y_\perp)}}\,,\ee
and does not involve the coupling constant at all.
Thus, this is formally like a free theory, although it has 
been obtained in a strong field regime where $g\alpha \sim 1$.
In fact, this is not really a free theory,
since the classical fields whose correlators are needed
are highly non-linear functionals of $g\alpha$, cf. eq.~(\ref{Atilde}).
However, as we shall shortly see, these classical
non-linear effects are not essential for getting saturation.

To determine the weight function $W_\tau[\alpha]$, one has to 
integrate the RGE (\ref{RGEH}) over all rapidities $\tau'$ with
$0\le \tau' \le \tau$. Clearly, the RPA leading to the Hamiltonian
(\ref{HLM}), which requires $k_\perp \ll Q_s(\tau')$,
is not appropriate for all such intermediate $\tau'$. % rapidities. 
Let $\bar\tau(k_\perp)$ be the rapidity at which the
saturation momentum becomes equal to the external momentum\footnote{We
are grateful to Al Mueller for pointing us the importance of the
separation scale $\bar\tau(k_\perp)$.} :
\be\label{barQs}
Q_s^2(\bar\tau(k_\perp))\,=\,k_\perp^2\,.\ee
Then eq.~(\ref{HLM}) applies at $\tau' > \bar\tau(k_\perp)$,
while for $0\le \tau' \le\bar\tau(k_\perp)$ we are rather in the
high-momentum regime. This suggests the following approximation:
\be\label{WHL}
W_\tau[\alpha]\,\approx\,W_{\bar\tau}^{\rm high}[\alpha]\,
W_{\bar\tau\rightarrow\tau}^{\rm low}[\alpha]\,,\ee
where $W_{\bar\tau}^{\rm high}[\alpha]$ is the weight
function in the high-momentum regime, eq.~(\ref{WHM}), 
evaluated at $\bar\tau(k_\perp)$ (this is the initial condition
for the evolution described by $H_{{\rm low}-k_\perp}$), and 
$W_{\bar\tau\rightarrow\tau}^{\rm low}[\alpha]$ is obtained
by integrating the RGE with Hamiltonian (\ref{HLM}) from
$\bar\tau(k_\perp)$ up to $\tau$.

A crude estimate for the separation scale $\bar\tau(k_\perp)$
can be obtained by using the solution in the high-momentum regime.
Eq.~(\ref{Qstau}) implies 
$Q^2_s(\tau)\propto {\rm e}^{\,c\bar\alpha_s\tau}$, where
$c=4$ in the DLA but will be kept here as a free parameter,
to partially account for our ignorance of the true dynamics at
$k_\perp \sim Q_s$. This, together with eq.~(\ref{barQs}), implies
\be\label{tbt}
\tau\,-\,\bar\tau(k_\perp)\,=\,{1\over c\bar\alpha_s}\,
\ln{Q_s^2(\tau)\over k_\perp^2}\,,\ee
which shows that, when $k_\perp \ll Q_s(\tau)$, we have also
$\tau\gg\bar\tau(k_\perp)$. Physically interesting correlation
functions like the gluon distribution (\ref{GCL}) involve generally an 
integral over all rapidities $0\le {\rm y}\le \tau$. In the regime
$\tau\gg\bar\tau(k_\perp)$, the dominant contributions to such
correlations come
from the interval $\bar\tau(k_\perp)\le {\rm y}\le \tau$, and are thus
determined by the piece 
$W_{\bar\tau\rightarrow\tau}^{\rm low}[\alpha]$
of the weight function (\ref{WHL}). Given the 
simplicity of the Hamiltonian (\ref{HLM}), this is easily obtained as
\cite{SATL} :
\be\label{WLM} 
W_{\bar\tau\rightarrow\tau}^{\rm low}[\alpha]
\,=\,{\cal N}_\tau\,{\rm exp}\left\{
-\,{1\over 2}\int_{\bar\tau(k_\perp)}^\infty
d{\rm y} \int {d^2 x_\perp}\
\partial^i \hat\alpha_{\rm y}^{\,a}(x_\perp)\,\zeta^{-1}_\tau({\rm y})
\,\partial^i \hat\alpha_{\rm y}^{\,a}(x_\perp)\right\},\ee
with the following width:
\be\label{ZLM}
\zeta_\tau({\rm y})\,=\,(1/\pi)\theta(\tau-{\rm y}).\ee
This implies (for $\bar\tau(k_\perp)\le {\rm y,\,y'}\le\tau$)
\be\label{alphaLM}
\langle \hat\alpha_{\rm y}^{\,a}(x_\perp)\,
\hat\alpha_{{\rm y}'}^{\,b}(y_\perp)\rangle_\tau&=&(1/\pi)
\delta^{ab}\delta({\rm y}-{\rm y}')\,
\langle x_\perp|\frac{1}{-\grad^2_\perp}|y_\perp\rangle,\\
\left\langle\partial^i\hat\alpha_{\rm y}^{\,a}(x_\perp)\,
\partial^i\hat\alpha_{{\rm y}'}^{\,b}(y_\perp)\right\rangle_\tau&=&(1/\pi)
\delta^{ab}\delta({\rm y}
-{\rm y}')\,\delta^{(2)}(x_\perp-y_\perp),
\label{FLM}\ee
showing that the probability 
distribution for the COV-gauge electric field
$\tilde {\cal F}^{+i}=-\partial^i\alpha$ is local and homogeneous
both in the transverse space and in space-time rapidity (within
the interval $\bar\tau(k_\perp)\le {\rm y} \le \tau$). This is consistent with
our assumption that transverse correlations in the system are over 
a typical scale $1/Q_s$, and thus cannot be resolved when the
hadron is probed with a much lower resolution. But this also
shows that the transverse $\delta$-function 
in eq.~(\ref{FLM}) must be taken with a grain of salt: this
is really delocalized over the correlation length $1/Q_s({\rm y})$.

It is now possible to compute the gluon distribution function
in this low momentum regime. Since the fields are strong,
we have to use the fully non-linear expression for the classical
field in eq.~(\ref{Atilde}). Eq.~(\ref{GCL})
involves the following 2-point function:
\be\label{FFLM}
\langle {\cal F}^{+i}_a(k^+,x_\perp) {\cal F}^{+i}_a
(-k^+,y_\perp) \rangle_\tau\,=\,
\int dx^- \int dy^- {\rm e}^{ik^+(x^--y^-)}
\langle {\cal F}^{+i}_a(x^-,x_\perp) {\cal F}^{+i}_a
(y^-,y_\perp) \rangle_\tau,\ee
which we compute as follows:
\be\label{FF0}
\langle {\cal F}^{+i}_a(\vec x) {\cal F}^{+i}_a
(\vec y) \rangle_\tau &=&
\left\langle\Bigl (U_{ab}^\dagger\partial^i\alpha^b\Bigr)_{\vec x}\,
\Bigl(U_{ac}^\dagger\partial^i\alpha^c\Bigr)_{\vec y}\right\rangle_\tau\nn
 &=&\left\langle \partial^i\alpha^b(\vec x)
\partial^i\alpha^c(\vec y)\right\rangle_\tau
\left\langle U_{ab}^\dagger 
(\vec x) U_{ca}(\vec y)\right\rangle_\tau\,,\ee
where, as indicated in the second line, the two COV-gauge
electric fields $\partial^i\alpha$ can be contracted only together
(mainly because of the path ordering in the Wilson lines,
which forbids other contractions \cite{JKMW97,KM98,SATL}).
By also using eq.~(\ref{FLM}), one obtains: 
\be\label{FFLM1}
\langle {\cal F}^{+i}_a(k^+,x_\perp) {\cal F}^{+i}_a
(-k^+,y_\perp) \rangle_\tau\,\approx\,{N^2-1\over \pi}\,
\int_{\bar\tau(k_\perp)}^\tau d{\rm y}\,\delta^{(2)}(x_\perp-y_\perp)
\,\left\langle U_{\rm y}^\dagger 
(x_\perp) U_{\rm y}(y_\perp)\right\rangle_\tau\,.\ee
The transverse $\delta$-function
in the r.h.s. of eq.~(\ref{FFLM1}) imposes $x_\perp = y_\perp$,
and therefore $\langle U_{\rm y}^\dagger U_{\rm y}\rangle 
\to 1$.
More precisely, this $\delta$-function is spread over a distance
$|x_\perp-y_\perp|\sim 1/Q_s({\rm y})$, which is precisely
the correlation length for the 2-point function
$\langle U_{\rm y}^\dagger 
(x_\perp) U_{\rm y}(y_\perp)\rangle$. Over such a distance,
the latter decreases by a factor $b$, with $b > 1$
but not much larger (typically, $b\sim e$). We thus obtain:
\be\label{FF}
\langle {\cal F}^{+i}_a(k^+,x_\perp) {\cal F}^{+i}_a
(-k^+,y_\perp) \rangle_\tau
\,= \,{N^2-1\over  \pi b}\,
\delta^{(2)}(x_\perp-y_\perp)\Bigl( \tau-\bar\tau(k_\perp)\Bigr). \ee
Up to the global factor $1/b$, this is the same result that would
have been obtained in the linearized, or weak field, approximation 
${\cal F}^{+i}_a \simeq -\partial^i\alpha_a$. All the non-linear
effects --- which were a priori encoded in the 2-point function
of the Wilson lines --- have dropped out from the gluon distribution
because of the locality of the propagator (\ref{FLM}).

By using eqs.~(\ref{GCL}) and (\ref{FF}), one can compute
the number of gluons per unit of transverse phase space in
this low-momentum regime. One obtains ($b_\perp$ is the impact
parameter) : %, $\int d^2b_\perp = \pi R^2$) :
\be\label{SAT}
\frac{d^2(x G )}{d^2k_\perp\,d^2b_\perp}\equiv
\int {d^2x_\perp\over 4\pi^3}\,{\rm e}^{-ik_\perp\cdot x_\perp}
\langle {\cal F}^{+i}_a(k^+,x_\perp) {\cal F}^{+i}_a
(-k^+,0_\perp) \rangle_\tau\,=\,{N^2-1\over 4 \pi^4}\,{1\over b}
\Bigl( \tau-\bar\tau(k_\perp)\Bigr),\, \ee
or, after also using eq.~(\ref{tbt}),
\be\label{WIGG}
\frac{d^2(x G )}{d^2k_\perp\,d^2b_\perp}\,
=\,{N^2-1\over 4 \pi^4 a}\,{1\over \bar\alpha_s}\,
\ln{Q_s^2(\tau)\over k_\perp^2}\,,\ee
where the two unknown constants $b$ and $c$ have been combined
into $a=bc$. 

Eq.~(\ref{WIGG}) is one of the main results in this paper.
It shows {\it marginal} saturation (when $k_\perp^2$ decreases,
the gluon density still increases, but only logarithmically), and is
consistent with the unitarity bounds: at fixed $k_\perp^2$,
the gluon density increases only linearly with $\tau$ (this is
most obvious on eq.~(\ref{SAT})), and so does also the
associated distribution function $xG(x,Q^2)$ for $Q^2\ll Q^2_s$.
The latter is easily obtained from eq.~(\ref{WIGG}) as
(with $\int d^2b_\perp = \pi R^2$):
\be\label{GDSAT}
xG(x,Q^2)&=&\frac{N^2-1}{4\pi^2 a}\,{1\over \bar\alpha_s}\,
R^2 Q^2\left[\ln(Q_s^2(\tau)/Q^2)+1\right]\nn
&=&\frac{N^2-1}{4\pi a N}\,{1\over \alpha_s}\,R^2 Q^2
\Bigl[c\bar\alpha_s(\tau\,-\,\bar\tau(Q)) +1\Bigr].\ee
If extrapolated up to $Q\sim Q_s$, this agrees
quite well with the corresponding extrapolation of the
gluon distribution at high momenta, as given by
eqs.~(\ref{GCLLIN}) and (\ref{Qsat}). 

According to eq.~(\ref{WIGG}), the gluon density at saturation
is of order $1/\alpha_s$, which corresponds to color fields as
strong as ${\cal F}^{+i}\sim 1/g$. Still, as the above analysis
clearly shows, the non-linear effects leading to saturation are not
those in the classical field, but rather those in the quantum
evolution, which led to the local weight function (\ref{WLM}).
In other terms, the saturation is built-in in the effective action
at low momenta. In this respect, our conclusions differ from those
in Refs. \cite{JKMW97,KM98}, although the final results found there
for the gluon density are formally similar to our eq.~(\ref{WIGG}).

On the other hand, our results here are consistent
with some recent analyses of the non-linear gluon evolution by
Mueller \cite{AM2} and Levin and Tuchin \cite{LT99}. 
In Ref. \cite{LT99}, Levin and Tuchin have obtained exact, semi-analytic,
solutions to the Balitsky-Kovchegov equation for
$\langle V^\dagger(x_\perp) V(y_\perp)\rangle_\tau$ \cite{B,K},
and then used these solutions to estimate the 
gluon distribution function. Their result is consistent with
%\be\label{LTxG}
%xG(x,Q^2)\,\propto\,R^2 Q^2\Bigl(4\bar\alpha_s\tau\,-\,
%\ln(Q^2/\Lambda_{QCD}^2)\,+\,{\rm const.}\Bigr),\ee
%where the ``constant'' depends upon the specific geometry
%of the hadron and the overall normalization is not under control.
%To compare with this, let us use eq.~(\ref{WIGG}) to compute the
%gluon distribution function within the present approach.
%This gives (for $Q^2\ll Q^2_s$):
%\be\label{GDSAT}
%xG(x,Q^2)&=&\frac{N^2-1}{4\pi^2 a}\,{1\over \bar\alpha_s}\,
%R^2 Q^2\left(\ln(Q_s^2(\tau)/Q^2)+1\right)\nn
%&=&\frac{N^2-1}{4\pi a N}\,{1\over \alpha_s}\,R^2 Q^2
%\Bigl(c\bar\alpha_s\tau\,-\,
%\ln(Q^2/\Lambda_{QCD}^2)\,+\,{\rm const.}\Bigr).\ee
%This is consistent with 
our eq.~(\ref{GDSAT}) provided one takes $c=4$, which coincides
with the prediction of the (a priori, crude) DLA (cf. eq.~(\ref{Qstau})).

As a check of the self-consistency of the RPA, let us compute
the 2-point function $\langle V^\dagger(x_\perp) V(0_\perp)\rangle_\tau$
for $x_\perp^2 = 1/k_\perp^2\gg 1/Q_s^2(\tau)$. This is again given
by eq.~(\ref{VV}) 
%namely \be
%\langle V^\dagger(x_\perp) V(0_\perp)\rangle_\tau&=&
%{\rm exp}\left\{-g^2 N\int_0^\tau d{\rm y}\,\Bigl[\gamma_{\rm y}
%(0_\perp)-\gamma_{\rm y}(x_\perp)\Bigr]\right\},\ee
where the integral over y is now decomposed into two pieces:
$0< {\rm y} < \bar\tau(x_\perp)$ and $\bar\tau(x_\perp)< {\rm y} <\tau$,
with $\bar\tau(x_\perp)\equiv \bar\tau(k_\perp^2=1/x_\perp^2)$
(cf. eq.~(\ref{barQs})). The first rapidity interval gives an
attenuation factor $b\sim e$, while in the second interval we can use
the propagator $\gamma_{\rm y}(x_\perp-y_\perp)$ from 
eq.~(\ref{alphaLM}). This gives:
\be\label{xioxLM}
\gamma_{\rm y}
(0_\perp)-\gamma_{\rm y}(x_\perp)\,=\,{1\over\pi}
\int 
{d^2p_\perp\over (2\pi)^2}\,\frac{1}{p_\perp^2}\,\Bigl[1-
{\rm e}^{ip_\perp\cdot x_\perp}\Bigr]\,\simeq\,{1\over 4\pi^2}
\,\ln\Bigl(Q_s^2({\rm y})x_\perp^2\Bigr),\ee
where the scale $Q_s({\rm y})$ has been 
introduced to cut off an ultraviolet divergence. 
%(Unlike in eq.~(\ref{xi1}), there is no infrared
%divergence in the integral in eq.~(\ref{xioxLM}).)
It can be shown  \cite{SATL}, via an analysis of the 
Balitsky-Kovchegov equation, that this is indeed the natural UV
cutoff introduced by non-linear effects. Since (cf. eq.~(\ref{tbt})):
\be
\ln(Q_s^2({\rm y})x_\perp^2)\,=\,c\bar\alpha_s({\rm y}
-\bar\tau(x_\perp)),\ee
we finally obtain:
\be\label{VVLM}
\langle V^\dagger(x_\perp) V(0_\perp)\rangle_\tau&\simeq&{1\over b}\,
{\rm exp}\left\{-c\bar\alpha_s^2\int_{\bar\tau(x_\perp)}^\tau d{\rm y}\,
({\rm y}-\bar\tau(x_\perp))\right\}\,\propto\,{\rm exp}\Bigl\{
-(c/2)\bar\alpha_s^2(\tau-\bar\tau(x_\perp))^2\Bigr\}\nn
&\propto&{\rm exp}\left\{-{1\over 2c}\Bigl[\ln
(Q_s^2(\tau)x_\perp^2)\Bigr]^2\right\}.\ee
Once again, this coincides with the corresponding result
in \cite{LT99} provided one takes $c=4$. Eq.~(\ref{VVLM})
shows that the correlator of the Wilson lines is rapidly
decreasing when $Q_s^2(\tau)x_\perp^2\gg 1$, so that the
RPA is indeed justified, at least as a mean field approximation.

Note finally that, in contrast to the weight function at high momenta,
eq.~(\ref{WHM}), the width of the Gaussian in eq.~(\ref{WLM}) 
does not involve any mass scale.  Thus, the weight function at low
momenta is scale invariant and universal 
(in the sense of being insensitive to the initial conditions,
and therefore the same for all hadrons).
These properties transmit to the correlation functions computed
with this weight function, up to logarithmic corrections which enter via
the lower rapidity limit $\bar\tau(k_\perp)$ (which depends logarithmically
on $k_\perp^2$) and via the ultraviolet cutoff $Q_s(\tau)$.
Thus, the correlation functions at saturation are functions of
$\ln(Q_s^2(\tau)/k_\perp^2)$, as manifest on 
eqs.~(\ref{WIGG}) and (\ref{VVLM}).

To conclude, in this paper we have constructed approximate solutions
to the RGE for the Color Glass Condensate valid at
either large, or small, $k_\perp$ momenta.
In both limits, we have used a mean field approximation. 
There is no difficulty of principle for using a mean field
approximation in more general cases, 
since this inevitably leads to a Gaussian form
for the functional $W_\tau[\alpha]$.  For strong fields, the width of this
Gaussian will satisfy a non-linear evolution equation,
which in the high-momentum regime can be shown \cite{SATL} to be consistent
with a previous result by Mueller and Qiu \cite{MQ}.
Beyond the mean field, we would like to be able to
understand the limitations of the mean field approximations and perhaps to
fully solve the renormalization group equation.
%\newpage

\bigskip
\bigskip
{\large{\bf Acknowledgements}}

One of us (E.I.) gratefully acknowledge illuminating discussions
with Al Mueller and Yuri Dokshitzer which allowed us to clarify some
aspects of the solution in the saturation regime. We equally thank
Ian Balitsky, Jean-Paul Blaizot,
Mikhail Braun and Eugene Levin for useful
conversations and/or correspondence.

This manuscript has been authorized under Contract No. DE-AC02-98H10886 
with the U. S. Department of Energy.

%\bigskip\bigskip

\end{document}